\documentclass[seceq]{ptptex}
\def\be{\begin{equation}}
\def\ee{\end{equation}}

\newcommand{\bea}{\begin{eqnarray}}
\newcommand{\eea}{\end{eqnarray}}

\title{
Toward a more realistic holographic QCD model\footnote{This talk is based on the paper \cite{HYY} .}
}

\author{
Mei \textsc{Huang}$^{1,2}$, Qi-shu \textsc{Yan}$^{3}$, Yi \textsc{Yang}$^{4}$%
}

\inst{
$^{1}$  Institute of High Energy Physics, Chinese Academy of Sciences, Beijing, China\\
$^{2}$  Theoretical Physics Center for Science Facilities, \\
Chinese Academy of Sciences, Beijing, China\\
$^{3}$  Department of Physics, National Tsing Hua University, Hsinchu, Taiwan\\
$^{4}$ Physics Division, National Center for Theoretical Sciences, Hsinchu,
Taiwan 
}

\abst{
By matching the $Dp-Dq$ system in type II superstring theory with the Regge trajectories of vector 
and axial-vector meson spectra, a possible 5-dimension metric of a $Dp-Dq$ soft-wall model has 
been determined.  The dependence of the metric parameters 
on Regge trajectory parameters has been demonstrated.  It is shown that the models 
defined in $Dp$ brane background for $p=3$ or $4$ can be consistent with experimental data.
}

\begin{document}

\maketitle

\section{Introduction}
The discovery of the gravity/gauge 
duality \cite{dual} has revived the hope to understand QCD in strongly coupled 
region using string theory. The gravity/gauge, or anti-de Sitter/conformal field theory 
(AdS/CFT) correspondence provides a revolutionary method to tackle the problem of 
strongly coupled gauge theories. For a review of AdS/CFT, see \cite{AdS/CFT} .
The string description of realistic QCD has not been successfully formulated
yet. Many efforts are invested in searching for such a realistic description
by using the ``top-down" approach, \textit{i.e.} by deriving holographic QCD
from string theory \cite{nonAdS-QCD}, as well as by using the ``bottom-up" approach, 
\textit{i.e.} by examining possible \textit{holographic} QCD models from experimental 
data \cite{Son:2003et,TB:05,DaRold2005,EKSS2005,Ghoroku:2005vt}.

It is an essential and crucial point for the realistic \textit{holographic} QCD model to 
reproduce Regge behavior. Regge behavior is a well-known feature of QCD \cite{Regge}, 
and it was the commanding evidence for suggesting the string-like structure of hadrons. 
A general empirical expression for Regge trajectories can be cast as
\begin{align}
M^{2}_{n,S}  &  = a_{n} \, n \, + \, a_{S} \, S + b \,, \label{ab}%
\end{align}
where $n$ and $S$ are the quantum number of high radial and spin excitations,
respectively. The slope $a_{n}$ and $a_S$ have dimension ${\rm GeV}^2$, and describe
the mass square increase rate in radial excitations and spin excitations, respectively. 
In principle, $a_{n}$ is not necessarily the same as
$a_{S}$, though $a_{n}=a_{S}$ can be taken as a good approximation by fitting
experimental data \cite{Anisovich:2000kx}. The parameter $b$ is the ground
state mass square, and it is channel-dependent.

Currently, one of the most successful models derived from string theory is Sakai-Sugimoto 
model \cite{SS}, which can describe spontaneously chiral symmetry breaking naturally, but 
fails to generate the linear Regge behavior. On the other hand, in the ``bottom-up" approach,  
many efforts have been paid to generate the linear Regge behavior for meson spectra 
\cite{KKSS2006,and06,kru05} and for 
baryon spectra \cite{Forkel}. In Ref. \cite{KKSS2006}, Karch, Katz, Son and Stephanov (KKSS) 
found that a $z^{2}$ correction in the dilaton field leads to the linear
behavior $M_{n}^{2} \propto n$. In the KKSS model, which is also called soft-wall model, 
the background metric structure are assumed to have only the logarithmic term in the  warp 
factor and the quadratic term in the dilaton field, the corresponding spectra for both vector and
axial vector mesons are given as $M_{n,S}^{2} = 4 n + 4 S$. Obviously, neither
$a_{n}$ nor $a_{S}$ can be consistent with experimental data, where they
should be close to one \cite{Anisovich:2000kx}. Moreover, in this simple
scenario, one cannot produce the ground state mass-square term $b$.

So a natural and important question is: what kind of \textit{holographic} model can
generate the realistic linear Regge behavior? We study this question by combining 
the ``top-down" method and ``bottum-up" method, \textit{i.e.} by matching $Dp-Dq$ system 
in type II superstring theory with the Regge trajectories of meson spectra in the hidden local 
symmetry model with group $SU(2)_L \times SU(2)_R$. 
We determine the possible structure of background metrics from the 
Regge trajectories of vector and axial-vector mesons. Our results show that the models defined in $Dp$ 
brane background for $p=3$ or $4$ can be consistent with experimental data
quite well. These results may shed lights 
on the correct prescription of the string theory dual to the realistic QCD 
and can be useful to construct a correct  phenomenological {\it holographic} model. 

\section{Regge trajectories of vector and axial-vector mesons}

We will consider the chiral symmetry breaking $SU(2)_L \times SU(2)_R \to SU(2)_V$ by using the 
Higgsless model, the Hidden local symmetry version of chiral symmetry breaking.  In this approach, 
the pseudo-scalar $\pi$ is the zero mode of axial-vector field and there is no its radial and higher 
spin excitations. Then we only need consider $S\geq 1$ mesons.
We take as samples the data of the radial and spin excitations of $\rho$ and 
$a$ from PDG2007 \cite{pdg2007}, which are listed in Table \ref{tablerhoa}. 
\begin{table}[th]
\begin{center}%
\begin{tabular}
[c]{|c|c|c|c|c|c|c|}\hline
n & $1$ & $2$ & $3$ & $4$ & $5$ & $6$\\\hline
$1^{--}_{\rho^{n}}$ & $0.770$ & $1.450$ & $1.700$ & $1.900$ & $2.150$ &
$2.270$\\\hline
$3^{--}_{\rho^{n}}$ & $1.690$ & $1.990$ & $2.250$ & $-$ & $-$ & $-$\\\hline
$1^{++}_{a_{1}^{n}}$ & $1.260$ & $1.640$ & $1.930$ & $2.095$ & $2.270$ &
$2.340$\\\hline
$3^{++}_{a_{3}^{n}}$ & $1.870$ & $2.070$ & $2.310$ & $-$ & $-$ & $-$\\\hline
\end{tabular}
\end{center}
\caption{Vector and axial-vector meson spectra (in GeV).}%
\label{tablerhoa}%
\end{table}
To describe Regge trajectories for both ($\rho_{1}$, $\rho_{3}$) and ($a_{1}$,
$a_{3}$), we use the general formula Eq. (\ref{ab}). From the experimental
data, the parameters of Regge trajectories can be determined by using the
standard $\chi^{2}$ fit. The parameters for $(\rho_1,\rho_3)$ mesons and their
correlations read
\begin{equation}%
\begin{array}
[c]{rl}%
a_{n}^{\rho}= \!\!\! & +0.91\pm0.23\\
a_{S}^{\rho}= \!\!\! & +1.08\pm0.39\\
b^{\rho}= \!\!\! & -1.09\pm1.18
\end{array}
\begin{pmatrix}
1 &  & \cr -0.82 & 1 & \cr +0.43 & -0.81 & 1
\end{pmatrix}
. \label{eq:rhoregge}%
\end{equation}
The parameters for $(a_1,a_3)$ mesons and their correlations read
\begin{equation}%
\begin{array}
[c]{rl}%
a_{n}^{a} = \!\!\! & +0.81\pm0.22\\
a_{S}^{a} = \!\!\! & +0.88\pm0.39\\
b^{a} = \!\!\! & +0.13\pm1.17
\end{array}
\begin{pmatrix}
1 &  & \cr -0.82 & 1 & \cr +0.43 & -0.81 & 1
\end{pmatrix}
. \label{eq:aregge}%
\end{equation}

It is known that chiral symmetry is spontaneously 
broken in the vacuum, thus the observed chiral partners are not degenerate. From the Regge
trajectories of the chiral partners $\rho$ and $a$, the chiral symmetry breaking in the 
vacuum is reflected by the difference of the ground state square-masses $b^{\rho}$ 
and $b^{a}$. The difference is as large as  $1~ {\rm GeV}^2$, \textit{i.e.} 
$b^{a}-b^{\rho}\simeq M_{a_1}^2-M_{\rho_1}^2 \simeq 1~{\rm GeV}^2$.

\section{Metric structure of $Dp-Dq$ system}

In order to investigate the possible dual string theory for describing Regge
behavior, we introduce the following $Dp-Dq$ branes system in type II
superstring theory. In the $Dp-Dq$ system, the $N_{c}$\ background $Dp$-brane
describe the effects of pure QCD theory, while the $N_{f}$ probe\ $Dq$-brane
is to accommodate the fundamental flavors which has been introduced by 
Karch and Katz \cite{Karch:2002sh}. 
Such a practice is well motivated from string theory side. For example, in the
Sakai-Sugimoto model, $p=4$ and $q=8$. Low energy hadronic excitations 
are fields on probe$Dq$ branes which is on the
background determined by the background $Dp$ branes.

First, we consider $N_{c}$\ background $Dp$-branes in type II superstring
theory. The near horizon solution in 10-dimension space-time is
\cite{Dp-brane}%
\begin{equation}
ds^{2}=h^{-\frac{1}{2}}\eta_{\mu\nu}dx^{\mu}dx^{\nu}+h^{\frac{1}{2}}\left(
du^{2}+u^{2}d\Omega_{8-p}^{2}\right)  ,\label{metric}%
\end{equation}
where\ $\mu,\nu=0,\cdots,p$, the warp factor $h\left(  u\right)  =\left(
R/u\right)  ^{7-p}$ and $R$ is a constant%
\begin{equation}
R=\left[  2^{5-p}\pi^{\left(  5-p\right)  /2}\Gamma\left(  \frac{7-p}%
{2}\right)  g_{s}N_{c}l_{s}^{7-p}\right]  ^{\frac{1}{7-p}}.
\end{equation}

The coordinates transformation (for the cases of $p\neq5$)%
$u=\left(  \frac{5-p}{2}\right)  ^{\frac{2}{p-5}}R^{\frac{p-7}{p-5}}z^{\frac
{2}{p-5}}$
brings the above solution (\ref{metric}) to the following \emph{Poincar\'{e}
form,}%
\begin{equation}
ds^{2}=e^{2A\left(  z\right)  }\left[  \eta_{\mu\nu}dx^{\mu}dx^{\nu}%
+dz^{2}+\frac{\left(  p-5\right)  ^{2}}{4}z^{2}d\Omega_{8-p}^{2}\right]  .
\end{equation}
Consider $N_{f}$\ probe $Dq$-branes with $q-4$ of their dimensions in
the $S^{q-4}$ part of $S^{8-p}$, 
the induced $q+1$ dimensions metric on the probe branes is given
as
\begin{equation}
ds^{2}=e^{2A}\left[  \eta_{\mu\nu}dx^{\mu}dx^{\nu}+dz^{2}+\frac{z^{2}}%
{z_{0}^{2}}d\Omega_{q-4}^{2}\right]  .\label{induced}%
\end{equation}
Where\ $\mu,\nu=0,\cdots,3$, and the metric function of the warp factor only
includes the logarithmic term
\begin{align}
A(z)=-a_{0} ~\mathrm{ln} z, ~~with ~~a_{0}=\frac{p-7}{2\left(  p-5\right) },
\end{align}
and the dilaton field part takes the form of
\begin{equation}
e^{\Phi} = \ g_{s}\left(  \frac{2}{5-p}\frac{R}{z}\right)  ^{\frac{\left(
p-3\right)  \left(  p-7\right)  }{2\left(  p-5\right)  }}.
\end{equation}
It follows that
\begin{equation}
\Phi(z) \sim d_{0} \ln z, ~with ~d_{0}=-\frac{\left(  p-3\right)  \left(
p-7\right)  }{2\left(  p-5\right)  }.\label{Phi}%
\end{equation}

The parameters $a_{0},d_{0}$ and other two parameters $k,c_{0}$ used in Sec.
\ref{sec-model} for any $Dp-Dq$ system are listed in Table \ref{c0d0}.
\begin{table}[th]
\begin{center}
\begin{tabular}
[c]{|c|c|c|c|c|c|c|c|c|}\hline
$p$ & \multicolumn{3}{|c|}{$3$} & \multicolumn{3}{|c|}{$4$} &
\multicolumn{2}{|c|}{$6$}\\\hline
$q$ & \multicolumn{2}{|c|}{$5$} & $7$ & $4$ & $6$ & $8$ & $4$ & $6$\\\hline
$k=-\frac{\left(  p-3\right)  \left(  q-5\right)  +4}{p-7}$ &
\multicolumn{3}{|c|}{$1$} & $1$ & $5/3$ & $7/3$ & $-1$ & $-7$\\\hline
$a_{0}=\frac{p-7}{2\left(  p-5\right)  }$ & \multicolumn{3}{|c|}{$1$} &
\multicolumn{3}{|c|}{$3/2$} & \multicolumn{2}{|c|}{$-1/2$}\\\hline
$c_{0}= k a_{0} $ & \multicolumn{3}{|c|}{$1$} & $3/2$ & $5/2$ & $7/2$ & $1/2$
& $7/2$\\\hline
$d_{0}=-\frac{\left(  p-3\right)  \left(  p-7\right)  }{2\left(  p-5\right)
}$ & \multicolumn{3}{|c|}{$0$} & \multicolumn{3}{|c|}{$-3/2$} &
\multicolumn{2}{|c|}{$3/2$}\\\hline
\end{tabular}
\end{center}
\caption{Theoretical results for the $Dp-Dq$ system.}%
\label{c0d0}%
\end{table}

We notice that $d_{0}=0$ for $D3$ background branes, \textit{i.e.} dilaton
field is constant in AdS$_{5}$ space. However, the dilaton field in a general
$Dp-Dq$ system can have a $\ln z$ term contribution, e.g. in the $D4-D8$
system (Sakai-Sugimoto model \cite{SS}), $d_{0}=-3/2$.

\section{The deformed $Dp-Dq$ soft-wall model}
\label{sec-model}

In the above section, we have derived the general metric structure of the $Dp-Dq$
system in Type II superstring theory, and we have noticed that the metric function
$A(z)$ only includes the logarithmic term, and in general there is another logarithmic 
contribution to the dilaton field.  However, from the lessons of AdS$_5$ metric ($D3$ system) 
and the Sakai-Sugimoto model($D4-D8$ system), the $Dp-Dq$ system cannot
describe linear trajectories of mesons.  It was shown in Ref.\cite{KKSS2006}, in order to 
produce linear trajectories, there should be a $z^2$ term, but all $z^2$ asymptotics should 
be kept in the dilaton field $\Phi(z)$ and not in the warp factor $A(z)$. Otherwise, the radial 
slope $a_n$ will be spin dependent.  Therefore, to describe the real QCD, we propose a 
deformed $Dp-Dq$ soft-wall model which is defined as
\begin{equation}
A(z)=-a_0 ~{\rm ln} z, \, ~\Phi(z)=d_{0}\ln z+c_{2}z^{2}\,.\label{scenario1}%
\end{equation}

By assuming that the gauge fields are independent of the internal space $S^{q-4}$,
after integrating out $S^{q-4}$, up to the quadratic terms and following the same assumption as in 
\cite{KKSS2006}, we can have the effective 5D action for higher spin mesons described by tensor 
fields as 
\begin{align}
I &  =\frac{1}{2}\int d^{5}x\sqrt{g}\,\,e^{-\Phi(z)}\bigg \{\Delta_{N}%
\phi_{M_{1}\cdots M_{S}}\Delta^{N}\phi^{M_{1}\cdots M_{S}}\nonumber\\
&  +m_{5}^{2}\phi_{M_{1}\cdots M_{S}}\phi^{M_{1}\cdots M_{S}}\bigg \},
\label{spin}
\end{align}
where $\phi_{M_{1}\cdots M_{S}}$ is the tensor field and $M_i$ is the tensor index. The value of $S$ 
is equal to the spin of the field. The parameter $m_{5}^2$ is the 5D mass square of the bulk fields, 
$g$ and $\Phi(z)$ are the induced $q+1$ dimension metric and dilaton field as 
shown in Eq. (\ref{induced}) and (\ref{Phi}). The action for $\rho_1, a_1$ 
and $\rho_3, a_3$ mesons is given by taking $S=1$, and $S=3$ respectively. 

Following the standard procedure of dimensional reduction with mode decompositions 
$\phi(x;z)=\sum_{n=0}\phi_n(x) \psi_n(z)$, the equation of motion (EOM) of bulk wavefunctions 
$\psi(z)_n$ for the general higher spin field can be derived
as
\bea
\partial_z^{2}\psi_{n} - \partial_z B \cdot \partial_z\psi_{n}
+\left(  M_{n,S}^{2} -m_{5}^{2}e^{2A}\right)  \psi_{n}
=0\,, \label{hispin}
\eea
where $M_{n,S}$ is the mass of the 4-dimension field $\phi_{n}(x)$ and%
\bea
B=\Phi-k (2\,S-1) A= \Phi+c_0 (2\,S-1) {\rm ln}z
\eea
is the linear combination of the metric background function and the dilaton field. It is worthy of 
remark that the spin parameter $S$ enters in the factor $B$ and can affect the EOM and spectra. 
The parameter $k$ is a parameter depending on the induced metric (\ref{induced}) of the $Dq$ brane. 
After integrating out $S^{q-4}$, $k$ is determined as $k=-\frac{\left(  p-3\right) \left(  q-5\right) +4}{7-p}$. 
It is obviously that $k$ depends 
on both $p$ and $q$. For simplicity, we have defined $c_0=k a_0= -\frac{\left(
p-3\right) \left(  q-5\right)  +4}{2\left( p-5\right) }$. 
The combination function $B(z)$ approaches logarithmic asymptotic at 
UV brane, and goes to $z^2$ asymptotic at IR region.

The parameters $c_0, d_0$ for any $Dp-Dq$ system are listed in Table  \ref{c0d0}.
In the following, we hope to determine the possible realistic \textit{holographic} QCD 
model from the Regge trajectories  of vector and axial-vector mesons. 

In the dictionary of AdS/CFT,  a $f-$form operator with conformal dimension 
$\Delta$ has 5-dimensional square-mass $m_5^2=(\Delta -f)(\Delta + f -4)$ in 
the bulk \cite{5Dmass},  and $m_{5}^{2}=0$ for both vector and axial-vector mesons. 
In the formalism of DBI, $m_5^2=0$ is also the consequence of gauge invariance.
Thus we take $m_{5,\rho}^{2}=m_{5,a}^2=0$ in the following calculation.
The spectra of EOM of Eq. (\ref{hispin}) has an exact solution:
\begin{equation}
M_{n,S}^{2}=4c_{2}n+4c_{2}c_{0}S+2c_{2}(1-c_{0}+d_{0})\,. \label{exact11}
\end{equation}
When $c_{0}=c_{2}=1$ and $d_{0}=0$, this solution reduces to results given in
the Ref. \cite{KKSS2006}. 
This exact solution strongly supports our parameterization on Regge trajectories and
can tell how phenomenological parameters $a_{n}$, $a_{S}$ and $b$ are directly
related with the metric parameters $c_{0}$, $c_{2}$ and $d_{0}$. This solution
shows that: 1) $a_{n}$ is completely determined by $c_{2}$, \textit{i.e.}
$c_{2}=a_{n}/4$; 2) $a_{S}$ is related with both $c_{0}$ and $c_{2}$, and it
is interesting to notice that $c_{0}=a_{S}/a_{n}$, which indicates that
$c_{0}$ reflects the difference of string tension in the radial direction and spin
direction; 3) The ground state square-mass $b$ is related with all metric parameters,
and $d_0$ can be solved out as $d_0=\frac{2 b}{a_n} + \frac{a_S}{a_n} - 1$.

If we take the approximation of $a_n=a_S=1$, we have $c_0=1, c_2=1/4$ for both
vector and axial-vector mesons, while $d_0$ is mainly determined by the ground state
square-mass as $d_0^{\rho/a}= 2b^{\rho/a}$. 

We use Eq. (\ref{exact11}) to fit the spectra of vector and axial-vector mesons. The
central values and correlation matrix for vector mesons read as
\begin{equation}%
\begin{array}
[c]{rl}%
c_{0}^{\rho}=\!\!\! & +1.19_{-0.39}^{+0.45}\\
c_{2}^{\rho}=\!\!\! & +0.23_{-0.06}^{+0.06}\\
d_{0}^{\rho}=\!\!\! & -2.24_{-1.68}^{+2.53}%
\end{array}%
\begin{pmatrix}
1 &  & \cr-0.39 & 1 & \cr +0.02 & -0.89 & 1
\end{pmatrix}
,\label{eq:rhoc0c2} 
\end{equation}
while the central values and correlation matrix for axial-vector mesons read as
\begin{equation}
\begin{array}
[c]{rl}%
c_{0}^{a}=\!\!\! & +1.09_{-0.44}^{+0.50}\\
c_{2}^{a}=\!\!\! & +0.20_{-0.06}^{+0.06}\\
d_{0}^{a}=\!\!\! & +0.51_{-2.36}^{+3.98}%
\end{array}%
\begin{pmatrix}
1 &  & \cr-0.43 & 1 & \cr +0.25 & -0.97 & 1
\end{pmatrix}
.\label{eq:aonec0c2}
\end{equation}

Since these two individual fitting suggest it is possible to use a common metric ( $c_0$ and $c_2$ ) 
to describe radial and higher spin excitations of both $\rho$ and $a_1$ mesons, we can fit 
the data in Table I with four theoretical free parameters $c_0$, $c_2$, $d_0^\rho$, 
and $d_0^a$, which read as
\begin{equation}
\begin{array}
[c]{rl}%
c_{0}=\!\!\! & +1.19_{-0.54}^{+0.54}\\
c_{2}=\!\!\! & +0.20_{-0.08}^{+0.06}\\
d_{0}^{\rho}=\!\!\! & -2.24_{-1.72}^{+4.51}\\
d_{0}^{a}=\!\!\! & +0.00_{-2.83}^{+4.20}%
\end{array}%
\begin{pmatrix}
1 &  & & \cr-0.93 & 1 & & \cr+0.88 & -0.98 & 1 & \cr +0.90 & -0.99 &+0.98 & 1
\end{pmatrix}
.\label{eq:cmb}
\end{equation}

The result shows that it is possible to economically 
accommodate Regge trajectories of both $\rho$ and $a_1$ with 
just $3$ nonvanishing parameters ($c_{0}$, $c_{2}$, and $d_{0}^{\rho}$).
Our fitting results show that $c_0$ prefers to the value of $1\sim1.5$, compared with
table \ref{c0d0} (For $D_3-Dq$ background brane case, both $q=5$ and $q=7$ probe brane
cases correspond to $c_0=1$, which is within the allowed overlapping region; 
for $D4$ background brane case, 
only $q=4$ case corresponds to
$c_0=1.5$ which is within the allowed overlapping region; 
for $D6$ background brane case, no $Dq$ probe 
brane case can be allowed in
the allowed overlapping region.), 
this indicates that the \textit{holographic} QCD model should be close 
to models defined in D$p$-branes background for $p=3$ or $4$. 

\section{Discussions and conclusion}

{\bf Discussion on $d_0$:}  In order to understand more about the logarithmic term in the dilaton
field, \textit{i.e.} $d_0 \ln z$, it should be helpful to compare our holographic model with an extended 
KKSS model, which only includes logarithmic term in $A(z)$ and only $z^2$ term in the dilaton 
field.  The extended KKSS model has the form of
\begin{equation}
A(z)=-a_0 ~{\rm ln} z, \, ~ \Phi(z)=c_{2}z^{2}\,.\label{scenario2}%
\end{equation}
By using the central values of Eqs. (\ref{eq:rhoregge}-\ref{eq:aregge}) 
and using the shooting method to find spectra from EOM and boundary conditions,
we determine the best value of $c_{0}$, $c_{2}$ and $m_5^2$ for radial and 
spin meson excitations, respectively.  
The best fitted values
of $(c_{0}^{\rho},c_{2}^{\rho},m_{5,\rho}^{2})$ and $(c_{0}^{a},c_{2}%
^{a},m_{5,a}^{2})$ are found to be
\begin{equation}%
\begin{array}
[c]{rl}%
c_{0}^{\rho}=\!\!\! & \,\,\,\,\,\,1.28,\\
c_{2}^{\rho}=\!\!\! & \,\,\,\,\,\,0.21\,\text{GeV}^{2},\\
m_{5,\rho}^{2}=\!\!\! & \,\,-0.06\,\text{GeV}^{2},
\end{array}%
\begin{array}
[c]{rl}%
c_{0}^{a}=\!\!\! & \,\,1.50,\\
c_{2}^{a}=\!\!\! & \,\,0.19\text{GeV}^{2},\\
m_{5,a}^{2}=\!\!\! & \,\,1.11\text{GeV}^{2}.
\end{array}
\end{equation}
In our fitting, we have taken the 5D mass as a free constant not a $z$ dependent function as 
treated in Refs.\cite{KKSS2006,Forkel}. It is noticed that the 5D square-mass for vector mesons
is almost zero, and is around $1 \text{GeV}^{2}$ for axial-vector mesons, which is a direct consequence 
of the chiral symmetry breaking $SU(2)_L \times SU(2)_R \rightarrow SU(2)_D$.  This is similar to the case 
in Ref.\cite{KKSS2006}, where 5D mass 
for $\rho$ is zero, and the axial-vector meson picks up a finite 5D mass via the Higgs mechanism
from the background scalar field.  
It is found that by using the extended KKSS model, in order to fit the vacuum square mass of
Regge trajectories, different nonzero 5D square-mass is required for vector and axial vector 
meson, respectively. It is suggested that $d_0^{\rho/a} {\rm ln}z$ contribution in the dilaton field  
in our holographic model proposed in Sec. IV
plays the same role as the $m_{5,\rho/a}^2$  in the extended 
KKSS scenario model, which might indicate that the chiral symmetry breaking is encoded in
the logarithmic term of the dilaton background. 

{\bf Discussion on $c_2$:} It is found that $c_2$ is not sensitive to
$Dp-Dq$ system, and mainly determined by the value of string tension or the slope 
of the radial excitations via
the relation $c_2=a_n/4$.  1) In \cite{KKSS2006}, $c_{2}$ was simply taken as $1~\mathrm{GeV}^{2}$.
 2) in the back  reaction holographic dual model\cite{Shock-2006}, $c_{2}$ is determined as $m_{q}^{2}/24$ 
 ($m_{q}$ the quark mass). For both cases, the $c_2$ cannot accommodate experimental data.
Andreev in Ref. \cite{Andreev-z2} shows that there is an upper bound of
$c_{2}$. $c_{2}$ can be determined by the coefficient $C_{2}$ \cite{c2} of the quadratic
correction to the vector-vector current correlator \cite{svz}
\begin{equation}
\label{PiVq2}\mathcal{N} q^{2}\,\frac{d \Pi_{V} }{dq^{2}}=C_{0}+\frac{1}%
{q^{2}}C_{2}+ \sum_{n\geq2}\frac{n}{q^{2n}}C_{2n}\langle\mathcal{O}%
_{2n}\rangle\,,
\end{equation}
where $C_{2}$ can be determined from $e^{+} e^{-}$ scattering data
\cite{Narison:1992ru}. According to \cite{Andreev-z2}, the relation between
$C_{2}$ in Eq. (\ref{PiVq2}) and the parameter $c_{2}$ for Dilaton field is:
$c_{2} = - \frac{3}{2} C_{2}$. The experimental bound $|\,C_{2}\,|\leq0.14
\,\text{GeV}^{\,2}$ gives $|\,c_{2}\,|\leq0.21 \,\text{GeV}^{\,2}$. Our
fitting results of $c_{2}$ is around $0.2$ and is within the experimental
upper bounds. 

In summary, the parameters of metric structure of the holographic QCD model is determined 
by matching $Dp-Dq$ soft-wall model in type II superstring theory with the Regge 
trajectories parameters of vector and axial-vector mesons spectra. Our results show that 
the models defined in $Dp$ brane background for $p=3$ or $4$ can be consistent with experimental 
data. According to the study of the 
reference \cite{Anisovich:2000kx},
the parameters $a_n$ and $a_s$ is almost universal 
for all mesons spectra due to their common strong interaction origin ( the subtle differences in $a_n$ 
and $a_s$ for different mesons can be attributed to higher order effectts or other dynamics like the 
chiral symmetry breaking ).  Although we have only used data of radial and higher spin exciations of 
$\rho$ and $a_1$ mesons,  the equation of motion of KK modes is 
universal to mesons with different CP charges, our conclusion should also be valid 
for other types of mesons.
We also show the dependence of the metric parameters on the Regge trajectories parameters: 
The quadratic term in the dilaton background field is solely determined by the slope in the radial direction;
The warp factor is mainly determined by the difference of the slope in the spin direction 
and the radial direction;  In order to produce the ground state square-masses, it is required to 
have different finite 5D square-mass or equivalently different logarithmic terms in the dilaton
 field for vector and axial-vector mesons, which encodes the information of the chiral 
 symmetry breaking.  This information is important for the construction of a realistic holographic QCD model.  
 Once the universal background metric  and meson-dependent parameters are determined from mass 
 spectra are determined from mass spectra, we can further test the $D_p-D_q$ soft-wall model by 
 calculating cross sections, decay widths, 
branching ratios, etc, of mesons.

\section*{Acknowledgments} M.H. thanks the organizers of the workshop "New Frontiers in QCD 2008".
The work of M.H. is supported by IHEP, and the CAS key 
project under grant No. KJCX3-SYW-N2, and NSFC under grant No. 10735040. The work 
of Q.Y. is supported by the NCS of Taiwan (No. NSC 95-2112-M-007-001
and 96-2628-M-007-002-MY3). The work of Y.Y.
is supported by NCTS through National Science Council of Taiwan.


\begin{thebibliography}{99}               

\bibitem {HYY}Mei Huang, Qi-shu Yan, Yi Yang,
  arXiv:0710.0988 [hep-ph].

\bibitem {dual}J.~M.~Maldacena,
Adv.\ Theor.\ Math.\ Phys.\ \textbf{2}, 231 (1998)
[Int.\ J.\ Theor.\ Phys.\ \textbf{38}, 1113 (1999)]; S.~S.~Gubser,
I.~R.~Klebanov and A.~M.~Polyakov,
Phys.\ Lett.\ B \textbf{428}, 105 (1998);
E. Witten,
Adv.Theor.Math.Phys. 2 (1998) 253-291.

\bibitem {AdS/CFT}O. Aharony, S.S. Gubser, J. Maldacena, H. Ooguri, Y. Oz,
\textit{Large N Field Theories, String Theory and Gravity, }Phys.Rept. 323
(2000) 183.

\bibitem {nonAdS-QCD}
O.~Aharony,
arXiv:hep-th/0212193;
A.~Zaffaroni,
PoS \textbf{RTN2005}, 005 (2005).

\bibitem {Son:2003et}D.~T.~Son and M.~A.~Stephanov,
Phys.\ Rev.\ D \textbf{69} (2004) 065020.

\bibitem {TB:05}
 G.~F.~de Teramond and S.~J.~Brodsky,
  Phys.\ Rev.\ Lett.\  {\bf 94}, 201601 (2005).

\bibitem {DaRold2005}L.~Da Rold and A.~Pomarol,
Nucl.\ Phys.\ B \textbf{721}, 79 (2005).

\bibitem {EKSS2005}J.~Erlich, E.~Katz, D.~T.~Son and M.~A.~Stephanov,
Phys.\ Rev.\ Lett.\ \textbf{95}, 261602 (2005).

\bibitem{Ghoroku:2005vt}
  K.~Ghoroku, N.~Maru, M.~Tachibana and M.~Yahiro,
  Phys.\ Lett.\  B {\bf 633}, 602 (2006)
  [arXiv:hep-ph/0510334].

\bibitem {Regge}G.~Veneziano,
Nuovo Cim.\ A \textbf{57}, 190 (1968);
P.D.B. Collins, \textit{An Introduction to Regge Theory and High Energy
Physics}, Cambridge Univ. Press, Cambridge (1975).

\bibitem {Anisovich:2000kx} A.V.~Anisovich, V.V.~Anisovich and
A.V.~Sarantsev,
Phys.\ Rev.\ D \textbf{62}, 051502 (2000).
  
\bibitem{SS}
 T.~Sakai and S.~Sugimoto,
  Prog.\ Theor.\ Phys.\  {\bf 113}, 843 (2005);
  Prog.\ Theor.\ Phys.\  {\bf 114}, 1083 (2006).
  
  
  \bibitem {KKSS2006}A.~Karch, E.~Katz, D.~T.~Son and M.~A.~Stephanov,
Phys.\ Rev.\ D \textbf{74}, 015005 (2006).

\bibitem{and06} O. Andreev, V.I. Zakharov, arXiv:hep-ph/0703010; 
Phys. Rev. D \textbf{74}, 025023 \ (2006).

\bibitem{kru05} M. Kruczenski, L. A. P. Zayas, J. Sonnenschein and D. Vaman, 
JHEP \textbf{06}, 046 (2005);  S. Kuperstein and J. Sonnenschein,
JHEP \textbf{11}, 026 (2004). 

\bibitem {Forkel}H.~Forkel, M.~Beyer and T.~Frederico,
JHEP \textbf{0707}, 077 (2007).
  
\bibitem {pdg2007}W.-M.Yao et al. (Particle Data Group), J. Phys. G 33, 1
(2006) and 2007 partial update for the 2008 edition.

\bibitem{Karch:2002sh}
  A.~Karch and E.~Katz,
  JHEP {\bf 0206}, 043 (2002)
  [arXiv:hep-th/0205236].
  
\bibitem {Dp-brane}G. W. Gibbons and K. Maeda,
\ Nucl. Phys. B298 (1988) 741; D. Garfinkle, G. T. Horowitz, and A.
Strominger,
\ Phys. Rev. D43 (1991) 3140; G. T. Horowitz and A. Strominger,
\ Nucl. Phys. B360 (1991) 197.


\bibitem{5Dmass}
E.~Witten,
  Adv.\ Theor.\ Math.\ Phys.\  {\bf 2}, 253 (1998).
  
\bibitem {Shock-2006}J.~P.~Shock, F.~Wu, Y.~L.~Wu and Z.~F.~Xie,
JHEP \textbf{0703}, 064 (2007).

\bibitem {Andreev-z2}O.~Andreev,
Phys.\ Rev.\ D \textbf{73}, 107901 (2006).

\bibitem{c2} L.~S.~Celenza and C.~M.~Shakin,
  Phys.\ Rev.\  D {\bf 34}, 1591 (1986).
  
 \bibitem {svz}M.A. Shifman, A.I. Vainstein, and V.I. Zakharov, Nucl.Phys.
B147, 385 (1979); 448 (1979).

  
\bibitem {Narison:1992ru}S.~Narison,
Phys.\ Lett.\ B \textbf{300}, 293 (1993).
  
\end{thebibliography}
\end{document}